\documentclass[aps,prd,showpacs,amsmath,nofootinbib,
superscriptaddress,onecolumn,preprintnumbers]{revtex4}
\usepackage{natbib}
\usepackage{amsmath}
\usepackage{amssymb}
\usepackage{graphicx}
\usepackage{color}
\usepackage{amsmath}
\usepackage{amsthm}
\usepackage{slashed}
\usepackage{graphicx}
\usepackage{soul}
\usepackage{cprotect}

\begin{document}

\title{Differing Manifestations 
of Spatial Curvature in Cosmological FRW Models}

\author{Meir Shimon}
\affiliation{School of Physics and Astronomy, 
Tel Aviv University, Tel Aviv 69978, Israel}
\email{meirs@tauex.tau.ac.il}
\author{Yoel Rephaeli}
\affiliation{School of Physics and Astronomy, 
Tel Aviv University, Tel Aviv 69978, Israel}
\affiliation{Department of Physics, University of California, San Diego, La Jolla, CA 92093, USA}
\email{yoelr@tauex.tau.ac.il}

\begin{abstract}
We find statistical evidence for a mismatch between the (global) spatial curvature parameter $K$ 
in the geodesic equation for incoming photons, and the corresponding parameter in the Friedmann equation that determines the time evolution 
of the background spacetime and its perturbations. The mismatch hereafter referred to as `curvature-slip' is especially evident when the SH0ES prior on the current expansion rate is assumed. 
This result is based on joint analyses of cosmic microwave background (CMB) observations with the PLANCK satellite (P18), 
first year results of the Dark Energy Survey (DES), 
Baryonic Oscillation (BAO) data, and 
- at a lower level of significance - also on 
Pantheon SNIa (SN) catalog. For example, the betting odds 
against the Null Hypothesis are greater than 
$10^7$:1, 1400:1 and 1000:1 when P18+SH0ES, P18+DES+SH0ES, and P18+BAO+SH0ES, respectively, are considered. Datasets involving SNIa weaken 
this curvature slip considerably. Notably, even when the SH0ES prior is not imposed the betting odds for the rejection of the Null Hypothesis are 70:1 and 160:1 in cases where P18+DES and P18+BAO are considered. When the SH0ES prior is imposed, global fit of the modified 
model (that allows for a nonvanishing `curvature slip') 
strongly outperforms that of $\Lambda$CDM as is manifested by significant Deviance Information 
Criterion (DIC) gains, ranging between 7 and 23, depending on the dataset combination considered. 
Even in comparison to K$\Lambda$CDM the proposed model results in significant, albeit smaller, 
DIC gains when SN data are excluded. Our finding could possibly be interpreted as an inherent 
inconsistency between the (idealized) maximally symmetric nature of the FRW metric, and the 
dynamical evolution of the GR-based homogeneous and isotropic $\Lambda$CDM model. As such it 
implies that there is an apparent tension between the metric curvature and the curvature-like term 
in the time evolution of redshift. 
\end{abstract}


\maketitle

\section{Introduction}

On the largest observable scales the Universe appears to be very isotropic around us. Augmented by 
the Cosmological Principle, this (`local') isotropy is endowed to every observer at rest in the cosmic 
microwave background (CMB) frame. The corresponding spacetime is uniquely described by the 
Friedmann-Robertson-Walker (FRW) metric. Space in the latter has either closed, flat or open geometry. 
If, in addition, it is assumed that gravitation is governed by General Relativity (GR), then this background spacetime is shaped by its homogeneous and isotropic matter content. 
Incoming photons propagating in curved space trace its global geometry, and observational 
determination of various (e.g. angular-diameter-, luminosity-distance, etc.) distance-redshift relations 
enable inference the matter content of the Universe. 
The CMB anisotropy and polarization depend on both the matter 
content and geometry.

The currently favored 
cosmological model, $\Lambda$CDM, is spatially flat, and its energy content 
consists of $\sim 69\%$ dark energy (DE), $31\%$ non-relativistic (NR) matter -- [$5\%$ baryons 
and $26\%$ cold dark matter (CDM)] -- as well as a residual amount $\lesssim 0.1\%$ CMB radiation, 
relativistic (and possibly also NR) neutrinos. The consensus `vanilla'  $\Lambda$CDM model, 
characterized by only six parameters, emerged from joint analyses of various cosmological datasets 
that include mainly CMB anisotropy and polarization, galaxy clustering and galaxy shear maps, type Ia supernovae luminosity distance measurements, 
and baryonic oscillations (BAO). 
However, these models are not fully 
consistent with each other. Perhaps the most glaring example is the disparity between the Planck 
satellite CMB data that favor a spatially closed cosmological model at $\sim 3\sigma$ statistical 
significance on the one hand, and BAO data that strongly favor flat space on the other hand, e.g. [1]-[4]. Their combination considerably weakens the case for a closed Universe, e.g. [5], resulting in the best-fit `model of choice' - the flat $\Lambda$CDM model. 

A spatially non-flat Universe would seem to be unnatural [6], i.e., fine-tuned, in either case of a 
curvature radius much larger or much smaller than the present horizon size of the observable Universe. 
For example, if the initial conditions of the Universe are set at the GUT scale, then the curvature radius 
must have been fine-tuned at a precision level of one part in a trillion so as to be either a quadrillion times larger or smaller than the horizon size of observable Universe at the present 
time. This is the consensus understanding of the `flatness problem' [6], although there are counter-arguments 
that lend some support to the claim that this is not really problematic for the original Hot Big Bang model 
[7]-[12]. This, as well as the `horizon problem',  
have provided the main motivation for the inflationary scenario. 
Observationally, whether space is flat or closed is still an unsettled issue, e.g. [1], [13]-[33], but the far reaching implications of a closed Universe cannot be overstated, e.g., [3].

Inflation addresses the flatness problem by positing a very early exponential 
expansion phase (sourced by vacuum-like energy) that ironed out any pre-existing 
spatial curvature and inhomogeneities, e.g. [34]. According to this scenario the scalar field that drives 
$\sim 60$ e-folds of exponential expansion eventually decayed and dissipated its energy to 
form ordinary matter, and even seeded the primordial density perturbations with a slightly red-tilted 
spectrum. It is hard to envision any residual curvature after such a violent expansion phase, unless 
inflation itself was so incredibly fine-tuned to have resulted in a spatially curved space after all.

Thus, from this specific theoretical perspective a spatially flat Universe is strongly favored 
(although open-, e.g. [35]-[38] and closed-inflation, e.g. [39] models 
were proposed); 
if so then the PLANCK dataset stands out, and is especially at odds with the
BAO dataset in favoring a non-flat Universe. The challenge is that only a probe of the very large 
scales can result in a credible inference of (or bounds on) the curvature radius. Perhaps the most precise and well-understood cosmological probe  
are measurements of 
CMB anisotropy and polarization, and so only satellite-based measurements are capable of meaningfully constraining the curvature radius; at present the Planck database is most optimally suited to do so. While joint analyses of the current CMB+BAO datasets tilt the balance towards a spatially flat space, it is 
arguably problematic to put too much weight on this conclusion, mainly because the current 
CMB+BAO datasets are internally inconsistent, e.g. [1]. Clearly, global geometry is a critical property of the Universe: whereas values of the other basic 
parameters of the cosmological standard model (SM) 
affect the (temporal) evolution of physical processes, only the curvature and DE parameters determine its global geometry and future evolution.

It is quite remarkable that GR, which has been directly probed (Equivalence Principle, 
time-variation of the gravitational constant, etc.) on relatively limited dynamical range -- from solar system down to sub-$\mu$m scales -- appears to be quantitatively valid also on galactic and cosmological scales. However, for GR to provide a reasonably good fit to all currently 
available observations, CDM and DE had to be invoked, and even then there are a few lingering 
anomalies in the GR-based cosmological model. Invoking a DE component that is 
$\sim 122$ orders of magnitude smaller than naively expected is perhaps one of the greatest puzzles in physics: It would be much more natural -- as was indeed commonly thought 
prior to its detection -- that DE identically vanishes due to some (yet unknown) symmetry principle; 
its cancellation to $\sim 122$ decimal places represents an incredible level of fine-tuning. In light of this realization, and as briefly explained in the present work, it could be equally likely that the inferred 
existence of DE-like component merely represents the breakdown of GR, rather than the existence of a genuine non-clustering vacuum-like contribution to the cosmic energy budget.

In this paper we explore certain hitherto unexplored extensions to the time-redshift, t(z), relation of  
the standard FRW-based treatment, and r(t) relation that in general departs from the geodesic equation. 
Basically, we allow for t(z) to differ from the relation obtained from the Friedmann equation, i.e. that 
does not depend solely on the matter content of the Universe. 
In other words, we employ t(z) relations in our analysis that represent a clear 
violation of local energy-momentum conservation, i.e. we basically 
abandon the GR-based framework. Energy-momentum non-conservation is a  generic feature of 
various theories of gravitation [40], most common (and well known) of 
which are scalar-tensor theories; other examples are Rastall gravity [41], unimodular gravity [42, 43], and other theories [44]-[47]. Perhaps unexpectedly, our analysis results point towards phenomenological parameters in the $r(z)$ relation that depart by up to $\sim 4\sigma$ from their canonical values 
depending on the dataset combination used and the assumed priors. This result adds to other anomalies/tensions that have been long recognized to afflict the cosmological SM, e.g. the `lensing anomaly', the $\sigma_{8}/S_{8}$ tension of the CMB with large scale clustering probes, and the `Hubble tension, 
e.g. [2], [31]. 

We employ the Deviance Information Criterion (DIC) 
in assessing the relative likelihood of each proposed model in comparison 
with both the standard flat $\Lambda$CDM and its `curved' extension K$\Lambda$CDM models. 
We find that depending on the specifics of dataset combinations the proposed models can be mildly 
or even strongly favored over $\Lambda$CDM, thereby substantiating our finding that the t(z) relation 
does not depend solely on the matter content from a global parameter analysis perspective.  

The paper is organized as follows. In section II we summarize a few basic results of the SM 
for reference in future sections and for setting the notation. In section III we describe our proposed modifications, and in section IV we outline the model extension explored here. Section V describes our model comparison analysis and results. Our conclusions are summarized in section VI. 
Throughout, we adopt units such that the speed of light, $c\equiv 1$.

\section{The Benchmark Standard Model}
 
On the largest observable scales the Universe is found to be highly 
isotropic to within one part in $10^{5}$. 
The implication of this observation is greatly widened by adoption of the Cosmological Principle, 
essentially that the Universe appears isotropic (and therefore, uniform) to every `fundamental' observer. The FRW metric 
of such a (maximally symmetric) space is expressed 
in terms of the line element
\begin{eqnarray}
ds^{2}=-dt^{2}+a^{2}\Big[\frac{dr^{2}}{1-Kr^{2}}+r^{2}(d\theta^{2}
+\sin^{2}\theta d\varphi^{2})\Big], 
\end{eqnarray}
where $a=a(t)$ is the scale factor, $K$ is the spatial curvature parameter, and (with no loss of generality) 
the origin of spatial coordinates is set to be at the observer. 
Incoming radial null geodesics in this spacetime integrate to
\begin{eqnarray}
r(\eta)= \left\{
\begin{array}{ll}
      \frac{\sin[\sqrt{K}(\eta_{0}-\eta)]}{\sqrt{K}} & ; K>0 \\
      \eta_{0}-\eta & ; K=0\\
      \frac{\sinh[\sqrt{-K}(\eta_{0}-\eta)]}{\sqrt{-K}} & ; K<0, \\
\end{array} 
\right. 
\end{eqnarray}
where $\eta\equiv\int\frac{dt}{a(t)}$ is the conformal time, and $\eta_{0}$ is its present value.

In the GR-based SM, use of the the FRW metric in the field equations for adiabatically 
evolving Universe can be summarized in a single Friedmann equation
\begin{eqnarray}
H^{2}+\frac{K}{a^{2}}=\frac{8\pi G\rho(a)}{3}, 
\end{eqnarray}
where $H\equiv\dot{a}/a$, $\dot{a}\equiv\frac{da}{dt}$, $G$ is the Universal 
gravitational constant, and $\rho$ is the {\it total} energy density.  
Assuming that the various contributions to the cosmic energy budget do 
not mutually interact, then the continuity equation applied to the i'th species,
\begin{eqnarray}
\dot{\rho}_{i}+3(1+w_{i})H\rho_{i}=0, 
\end{eqnarray}
where $w_{i}$ is the equation of state (EOS) parameter  
of the i'th species, integrates to
\begin{eqnarray}
\rho_{i}(a)=\rho_{i,0}\exp\Big[-\int 3(1+w_{i}(a))\frac{da}{a}\Big]. 
\end{eqnarray}
In addition, if it is assumed 
that $w_{i}(a)=w_{i}$ is fixed 
(an assumption that breaks down in the case of neutrinos when they turn 
NR once the universe cools down below a temperature equivalent to 
their rest masses), an assumption that is {\it not} made in actual computations, 
then $\rho_{i}a=\rho_{i,0}a^{-3(1+w_{i})}$, and the total energy density is 
\begin{eqnarray}
\rho(a)=\sum_{i}\rho_{i,0}a^{-3(1+w_{i})}. 
\end{eqnarray}

Substitution of Eq. (6) in Eq. (3) and integration yield the basic expression for $t(a)$, or equivalently $\eta(z)$, where $a\equiv (1+z)^{-1}$. Use of the latter expressions in Eq. (2) results in the following $r(z)$ relation
\begin{eqnarray}
H_{0}r(z)= \left\{
\begin{array}{ll}
      \frac{\sin(\sqrt{-\Omega_{k}}\mathcal{D})}{\sqrt{-\Omega_{k}}} & ; \Omega_{k}<0, K>0 \\
      \mathcal{D} & ; \Omega_{k}=0, K=0\\
      \frac{\sinh(\sqrt{\Omega_{k}}\mathcal{D})}{\sqrt{\Omega_{k}}} & ; \Omega_{k}>0, K<0, \\
\end{array} 
\right.
\end{eqnarray}
where $\Omega_{k}\equiv -K/H_{0}^{2}$ and
\begin{eqnarray}
\mathcal{D}(z;\{\Omega_{i,0}\})&\equiv &H_{0}(\eta_{0}-\eta)=\int_{0}^{z}\frac{dz'}{E(z')},\nonumber\\
E^{2}(z')&\equiv &\sum_{i}\Omega_{i,0}(1+z')^{3(1+w_{i})}.
\end{eqnarray} $\{\Omega_{i,0}\}$ collectively denotes the various $\Omega_{i,0}$ which are the respective values of the energy densities  at present, expressed in 
critical density units, $\rho_{c}\equiv 3H_{0}^{2}/(8\pi G)$, including $\Omega_{k}$ 
with an effective EOS $w_{k,eff}=-1/3$. 
Specifically, in a Universe containing only photons, dust, DE, and curvature
\begin{eqnarray}
E(z)\equiv\sqrt{\Omega_{r}(1+z)^{4}+\Omega_{m}(1+z)^{3}+\Omega_{k}(1+z)^{2}+\Omega_{\Lambda}}.
\end{eqnarray}

It should be noted that $\Omega_{k}$ controls both the {\it dynamics} of cosmic evolution (via 
the Friedmann equation, Eq. 3, or equivalently Eq. 9) and the {\it kinematics}
(or perhaps more precisely referred to as the `optics') of incoming photons (via the geodesic equation, Eq. 2). 
Also important to keep in mind that whereas the optics only depends on the 
the form of the metric, essentially reflecting the spatial symmetry, the dynamics 
depends on the underlying theory of gravitation (that relates geometry to 
the undrlying energy-momentum content) as well.

Given certain initial conditions, i.e. the density and velocity perturbation fields at $\eta=0$, then 
any individual Fourier mode evolves independently of all the other modes in the linear regime. 
Perturbation levels at recombination and shortly after are at the $O(10^{-5})$ and the linear approximation 
is reasonably good. The evolution histories of photons and neutrinos are governed by the collisional and collisionless Boltzmann equations, respectively. Energy density perturbations, the velocity field and gravitational potential satisfy the coupled continuity, Euler and perturbed Einstein equations on the FRW background. All these equations are mutually coupled. The primordial gravitational potential is characterized by a power spectrum $P(k)$, and the latter evolution of density perturbations at any given Fourier mode is characterized by the transfer function 
$T(k;\eta)$. Three-dimensional density perturbations, velocity, and metric perturbations are 
Fourier-expanded in terms of the eigenfunctions of the 3D laplacian. In case of flat space these are the 
3D planewaves that, in turn, are separated via Rayleigh expansion into their angular and radial dependencies. 
The former is represented in terms of the spherical harmonics and the latter in terms of spherical Bessel functions. In the case of open or closed space the latter are replaced by the `ultraspherical' Bessel functions, 
e.g. [48]-[52]. The observed {\it angular} power-spectra, bispectra, tri-spectra, etc., of the CMB anisotropy 
and polarization result from projection on the 2D celestial sky. The radial distance as a function of time/redshift that appears in either the spherical- or ultraspherical-Bessel functions is given by Eq. (2). In general, the global geometry of space is manifested on the largest angular scales, but spatial curvature 
in particular leaves imprints also on general features of the power spectrum, e.g. [53].

\section{Modified Curvature SM}

The primary motivation for the present work comes from the specific way by which the cosmological 
constant is included in the energy budget of the SM: In the model the energy budget and its perturbations govern the background field 
and linear perturbation equations, respectively, 
i.e. $G_{\mu}^{\nu}=8\pi G T_{\mu}^{\nu}$ and $\delta G_{\mu}^{\nu}=8\pi G \delta T_{\mu}^{\nu}$.
The energy-momentum tensor associated with the cosmological constant is $T_{\mu}^{\nu}=
\delta_{\mu}^{\nu}\Lambda$, where $\Lambda$ is the spacetime-independent cosmological constant. 
This implies that the corresponding $\delta T_{\mu}^{\nu}$ vanishes, i.e. this simplest type of DE simply 
does not cluster and leaves no perturbation imprints. 
How is this any different from simply putting `by hand' in Eq. (9) a parameter $\Omega_{\Lambda}$ 
that has nothing to do with energy density, but rather a dimensionless parameter $\Omega_{\Lambda}$ 
with a trivial redshift-dependence that is introduced 
{\it ad hoc}? 
Such a term would represent the breakdown of local energy-momentum conservation, i.e. 
$T^{\mu}_{\nu;\mu}=0$, where `;' denotes covariant derivative, 
and with it the breakdown of GR, at least on cosmological scales. Put alternatively, 
had we assumed that the Universe contains all the usual matter ingredients but no DE, then the 
data analysis would have resulted in an `anomaly parameter' with the value of 0.69, drastically 
different from its expected vanishingly small value. This could have then 
been legitimately viewed as the breakdown of GR on cosmological scales at very high confidence level. As a side benefit, this view would imply that the vacuum energy on cosmological scales exactly vanishes, as thought to be the case before the discovery of the accelerated expansion. The alternative of insisting on the validity of GR on cosmological scales comes at a hefty $\sim 122$ orders of magnitude tension between the GR-based SM of cosmology and quantum field theory (QFT).

Specifically in the present work, we explore implications of phenomenologically modifying 
the time-redshift, $t(z)$, 
and distance-time, $r(t)$, relations on cosmological scales  
(on which the assumption of homogeneity and isotropy is still assumed). 
The SM $t(z)$ relation follows from the Friedmann equation, i.e. local energy-momentum conservation. 
In addition, the $r(t)$ relation describes the trajectories of incoming radiation along null-geodesics. 
The latter describe kinematics of test particles under the premise of energy-momentum conservation. Combining r(t) and t(z) yields the observable r(z) relation.

Here we replace Eq. (7) with
\begin{eqnarray}
H_{0}r(z)= \left\{
\begin{array}{ll}
      \frac{\sin(\sqrt{-\omega_{k}}\mathcal{D})}{\sqrt{-\omega_{k}}} & ; \omega_{k}<0\\
      \mathcal{D} & ; \omega_{k}=0\\
      \frac{\sinh(\sqrt{\omega_{k}}\mathcal{D})}{\sqrt{\omega_{k}}} & ; \omega_{k}>0\\
\end{array} 
\right.
\end{eqnarray}
where we introduced a new parameter $\omega_{k}$ while leaving $\mathcal{D}$, Eq. (8), unchanged.
In general, we do {\it not} require $\omega_{k}$ to be equal to  $\Omega_{k}$ in Eqs. (9) and (10), and define $\kappa\equiv\omega_{k}-\Omega_{k}$ as a derived parameter in our modified models. 
Clearly, nested within these models are K$\Lambda$CDM, if indeed 
$\omega_{k}=\Omega_{k}$, and flat $\Lambda$CDM in the particular case 
$\omega_{k}=\Omega_{k}=0$. In a similar vein to the addition of the cosmological constant to the 
SM as discussed above, we introduce the free dimensionless parameter $\kappa$ coupled to a specific 
redshift dependence (as shown in Eqs. 9 and 10) but with neither energetic interpretation nor any 
perturbation imprints, i.e. it does not cluster, much like DE (assuming it is a cosmological constant). Obviously, the likelihood of $\kappa \neq 0$ will be determined by contrasting the modified model with the various datasets. Significant preference of non-zero values could potentially signal 
inadequacy of GR, first indication for which could have been the very deduction that $\Lambda\neq 0$ 
some 25 years ago. 

Because no perturbations are considered in the value of $\Omega_{k}$,  in our analysis $\kappa$ is a truly anomaly parameter essentially devoid of physical meaning. In this specific 
sense it plays a role similar to other phenomenological parameters that have been considered in various generalizations of the SM, such as the lensing anomaly parameter, $A_{L}$ e.g. [54], the dipole and integrated Sachs Wolfe (ISW) anomaly parameters, e.g. [55, 56], and the CMB temperature mismatch parameter when 
local priors on $H_{0}$ are assumed [57]-[59]. The parameter $\kappa$ should be consistent with zero 
if GR and the underlying assumption of the Cosmological Principle 
are not just qualitatively but also quantitatively valid. 
In principle, since $\Omega_{k}$ appears in Eq. (9) as a contribution of an effective fluid with equations of state $w=-1/3$, 
it could be viewed as the combined effect of curvature and a `K-matter contribution' (save for the fact that it is unperturbed in our analysis), e.g. [60]. However, as we will see below, it turns out that 
our analysis of the (present) observational datasets yields $\kappa>0$, further reducing the likelihood for it to be a new exotic matter contribution candidate. In general terms, the `null hypothesis' is 
rejected to the extent that $\kappa$ statistically departs from zero. 

One possible explanation for the nonvanishing phenomenological 
parameter $\kappa$ could be that the FRW spacetime metric is an over-idealization 
of spacetime that fails on sufficiently small scales. 
This could possibly indicate that averaging over 
sufficiently small scales second-order perturbations do not vanish and could back-react on the smoothed-out 
background. In particular, it can manifest in the form of effective spatial curvature or K-matter, DE, and 
also stress-like contributions in the volume-averaged Friedmann equation. This does not readily and unequivocally reflect 
on the effective spacetime metric. These effects could not only 
tilt the balance between the various terms in the volume-averaged 
Friedmann equation, but 
can also affect the growth of structure, thereby biasing the inferred cosmological parameters, 
the curvature parameter included, e.g. [61]-[76].

\subsection{Models and Datasets}

We have carried out an extensive statistical analysis in order to determine the most probable ranges 
of the extended parameter set of the proposed modifications to the SM. To the `vanilla' {\it fundamental} 
model parameters $\Omega_{b}h^{2}$, $\Omega_{c}h^{2}$, $\tau$, $A_{s}$ and $n_{s}$ we add both 
$\Omega_{k}$ and $\omega_{k}$. The parameter $\kappa$ is a derived parameter. Additional relevant derived parameters that will be of interest to us here are $H_{0}$, $S_{8}$ and $\sigma_{8}$, especially 
in light of the well-known tensions involving these parameters when inferred from various dataset 
combinations. The models considered in this work are summarized in Table I.

\begin{table}[h]
\begin{tabular} {|c |c|}
\hline 
\hline
 \verb|Model 1| ($\Lambda$CDM) & $\Omega_{k}=0=\omega_{k}$\\
\hline
 \verb|Model 2| (K$\Lambda$CDM) & $\Omega_{k}\equiv\omega_{k}$\\
\hline
 \verb|Model 3| \ \ \ \ \ & $\Omega_{k}\neq 0$, $\omega_{k}\neq 0$\\
\hline
 \verb|Model 4| \ \ \ \ \ & $\Omega_{k}\neq 0$\\
\hline
 \verb|Model 5| \ \ \ \ \ & $\omega_{k}\neq 0$\\
\hline
\end{tabular}
\caption{Models explored in this work.} 
\end{table}
 
The datasets employed in this analysis constitute the standard commonly adopted set (in model testing) 
which is included (along with their corresponding likelihood functions) as part of the 
CosmoMC 2021 package. These include the CMB Planck 2018 data [77], DES 1 yr [78], BAO (data compilation from BOSS DR12 [79], MGS [80], and 6DF [81]), and Pantheon data (catalog of 1048 
SNIa in the redshift range $z\lesssim 2$ [82]). The entire Planck dataset is included, with multipole 
range $2<l<2500$ covered by the \verb|plikHM_TTTEEE|, \verb|lowl| and \verb|lowE| likelihood 
functions.

As in our previous (somewhat) related work [59], we consider P18, P18+DES, P18+BAO, P18+SN and P18+BAO+SN, dataset 
combinations, for each of which we performed the analysis 
with and without the SH0ES prior on $H_{0}$. Sampling from posterior distributions is done 
using the fast-slow dragging algorithm with a Gelman-Rubin convergence criterion 
$R-1<0.02$ (where R is the scale reduction factor). 
Testing the the curved space model (with the new parameters included) using only the Planck dataset (without the SH0ES prior) did not satisfy this convergence criterion; 
therefore, results of this case are not considered in this work.

\subsection{Results}

For each of the five models described above we compute the DIC for 
each dataset combination. The DIC is defined as [83]
\begin{eqnarray}
DIC\equiv 2\bar{\chi^{2}}(\theta)-\chi^{2}(\bar{\theta}), 
\end{eqnarray}
where $\theta$ is the vector of free model parameters and bars denote averages over the posterior distribution $P(\theta)$. By ‘Jeffreys scale’ convention, a model characterized by 
$\Delta$DIC that is lower with respect to a reference model by $<$1, 1.0-2.5, 2.5-5.0, and $>$5, would be considered 
as inconclusively, weakly/moderately, moderately/strongly, or decisively favored [83], respectively, over 
the reference model. These likelihood categories are summarized in Table II.

Our quantitative DIC results are summarized in Tables III and IV. 
Specifically, considering dataset combinations 
with the SH0ES prior, fits to \verb|Model 3| yield DIC values that are lower by 7-23 in comparison with 
\verb|Model 1|. When the SH0ES prior is excluded, \verb|Model 3| is significantly favored over  for the 
dataset combinations P18+DES, and P18+SN, with DIC gains of 8 and 7, respectively, but does not outperform the \verb|Model 1| fits with the datasets combinations P18+BAO, and P18+SN+BAO. 
Even when compared to \verb|Model 2|, the \verb|Model 3| fits are better when the SH0ES prior is included. When  the SH0ES prior is excluded, \verb|Model 1| and \verb|Model 2| yield quite 
similar DIC values (except in the case of P18 data alone when \verb|Model 2| is strongly favored over \verb|Model 1| as is shown in Table III); therefore, \verb|Model 3| improves over \verb|Model 2| at about the same extent that it improves over \verb|Model 1|. 
\verb|Model 3| is found to be favored over \verb|Model 4| and \verb|Model 5| in fits 
to all dataset combinations with the SH0ES prior. 

\begin{table}[h]
\begin{tabular} {|c | c | c|c|c|}
\hline 
\hline
$\Delta$ DIC & $<1$ & $1.0-2.5$ & $2.5-5.0$ & $>5.0$\\
\hline
\hline
Evidence & inconclusive & weak/moderate & moderate/strong & decisive \\
\hline
\end{tabular}
\caption{Jeffrey scale.}
\end{table}

\begin{table}[h]
\begin{tabular} {|c | c | c|c|c|c|}
\hline 
\hline
Datasets & \verb|Model 1| & \verb|Model 2| & \verb|Model 3| & \verb|Model 4| & \verb|Model 5|\\
\hline
\hline
P18       & 2799.4 & 2791.0 & -- & 2800.1 & 2791.3\\
\hline
      & & -8.3 & -- & 0.7 & -8.1\\
\hline
P18+DES   & 3339.8 & 3338.4 & 3331.6 & 3338.4 & 3339.2\\
\hline
      & & -1.3 & -8.1 & -1.3 & -0.6\\
\hline
P18+BAO   & 2805.9 & 2806.5 & 2805.9 & 2806.8 & 2806.4\\
\hline
      & & 0.6 & 0 & 0.9 & 0.5\\
\hline
P18+SN    & 3834.8 & 3835.0 & 3827.7 & 3835.9 & 3835.7\\
\hline
      & & 0.2 & -7.0 & 1.1 & 0.9\\
\hline
P18+SN+BAO& 3840.8 & 3841.3 & 3841.7 & 3841.3 & 3841.4\\
\hline
      & & 0.5 & 0.9 & 0.5 & 0.6\\
\hline
\hline
\end{tabular}
\cprotect\caption{DIC values for the various models and dataset combinations. 
The reference \verb|Model 1| and \verb|Model 2| are the flat $\Lambda$CDM 
and K$\Lambda$CDM models, respectively.} 
\end{table}

\begin{table}[h]
\begin{tabular} {| c | c|c|c|c|c|c|c|c|c|c|c|}
\hline 
\hline
Datasets & \verb|Model 1| & \verb|Model 2| & \verb|Model 3| & \verb|Model 4| & \verb|Model 5|\\
\hline
\hline
P18+SH0ES  & 2819.3 & 2809.3 & 2795.7 & 2804.8 & 2807.8 \\
\hline
      & & -10.0 & -23.6 & -14.5 & -11.6\\
\hline
P18+DES+SH0ES   & 3354.6 & 3342.0 & 3331.9 & 3338.4 & 3340.1\\
\hline
      & & -12.6 & -22.7& -16.2 & -14.5\\
\hline
P18+BAO+SH0ES   & 2824.8 & 2821.5 & 2812.2 & 2818.4 & 2820.9\\
\hline
      & & -3.3 & -12.6& -6.4 & -3.9\\
\hline
P18+SN+SH0ES   & 3853.8 & 3844.9 & 3843.1 & 3845.0 & 3844.2\\
\hline
      & & -8.9 & -10.7& -8.8 & -9.6\\
\hline
P18+SN+BAO+SH0ES& 3859.4 & 3856.8 & 3852.4 & 3854.3 & 3856.0\\
\hline
      & & -2.6 & -7.0& -5.2 & -3.4\\
\hline
\hline
\end{tabular}
\caption{
As in Table III, but with the SH0ES prior is included.} 
\end{table}

\begin{figure}[h]
\begin{center}
\leavevmode
\includegraphics[width=0.65\textwidth]{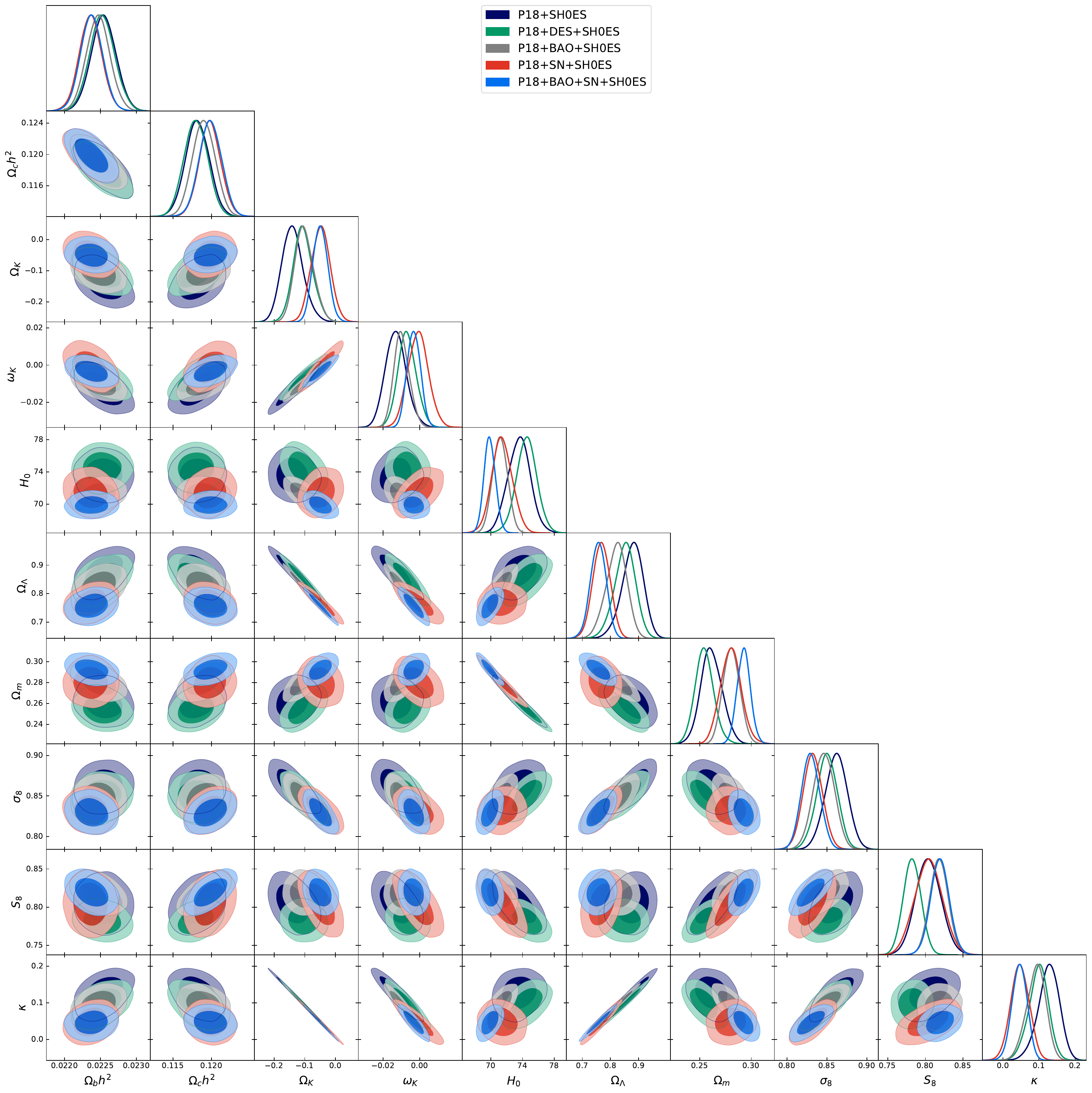}
\includegraphics[width=0.65\textwidth]{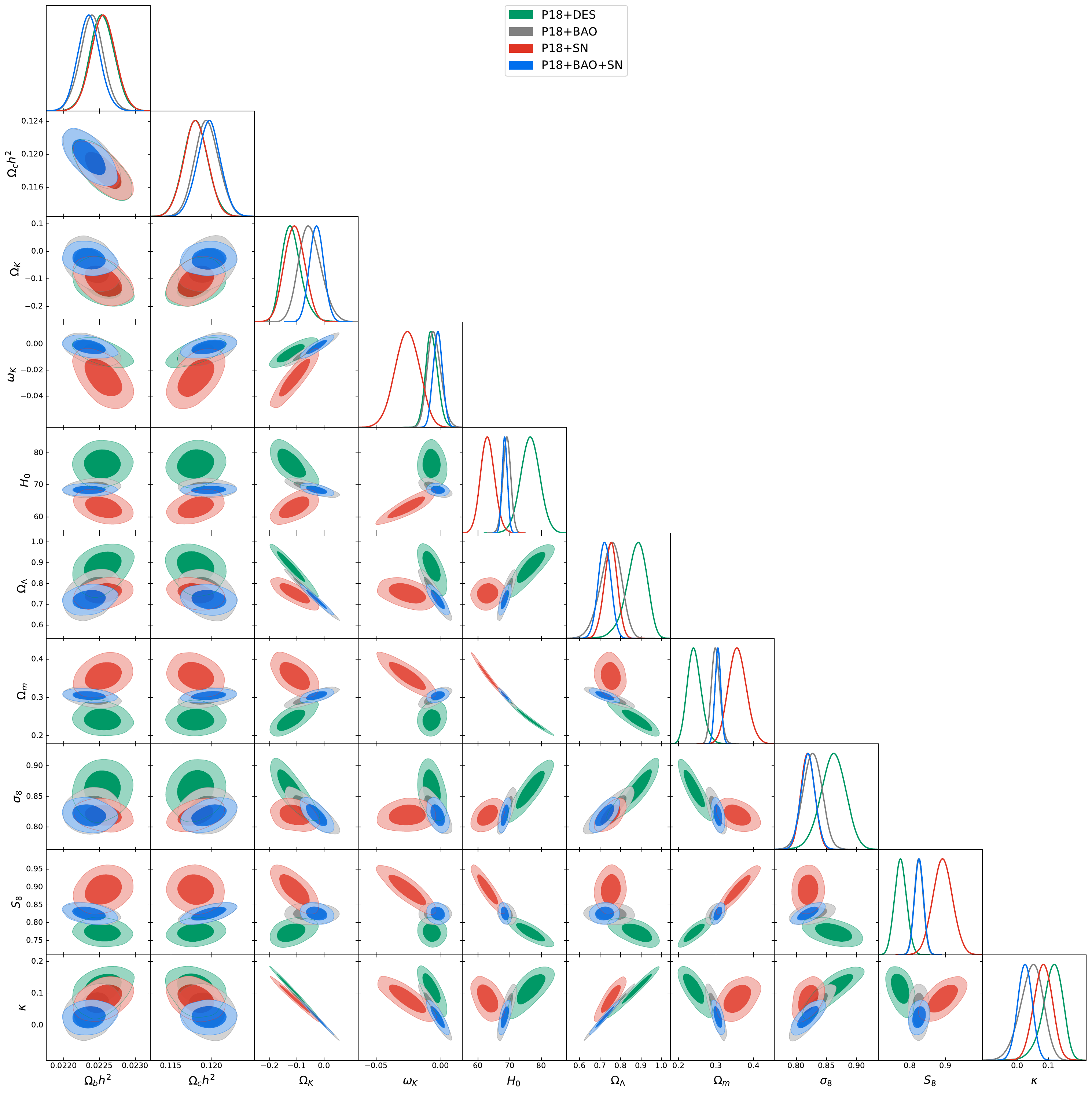}
\end{center}
\cprotect\caption{
Confidence contours and posterior parameter distributions in   
\verb|Model 3|: $\Omega_{k}\neq 0$ and $\omega_{k}\neq 0$.}
\end{figure}

\begin{table}[h]
\begin{tabular} {|c | c | c | c |}
\hline 
\hline
Dataset & $\Omega_{k}$ & $\omega_{k}$ & $\kappa$\\
\hline
\hline
P18+SH0ES &  1:500 & 1:20 & 1:10000000\\
\hline
P18+DES+SH0ES &  1:300 & 1:5 & 1:1400\\
\hline
P18+BAO+SH0ES &  1:1250 & 1:25 & 1:1000\\
\hline
P18+SN+SH0ES &  1:7 & {\rm fair} & 1:14\\
\hline
P18+BAO+SN+SH0ES & 1:20 & {\rm fair} & 1:25\\
\hline
P18 &  -- & -- &  --\\
\hline
P18+DES &  1:50 & 1:7 & 1:70\\
\hline
P18+BAO &  1:4 & {\rm fair} & 1:5\\
\hline
P18+SN &  1:200 & 1:200 & 1:160\\
\hline
P18+BAO+SN & {\rm fair} & {\rm fair} & {\rm fair}\\
\hline
\end{tabular}
\cprotect\caption{Betting odds against $\Omega_{k}\neq 0$, $\omega_{k}\neq 0$ and $\kappa\neq 0$ in 
\verb|Model 3| for various dataset combinations. Whereas either $\Omega_{k}\neq 0$ or 
$\omega_{k}\neq 0$ rule out flat $\Lambda$CDM, $\kappa\neq 0$ rules out both flat $\Lambda$CDM 
and K$\Lambda$CDM.}
\end{table}

\begin{table}[h]
\begin{tabular} {|c | c | c|c|}
\hline 
\hline
Dataset & \verb|Model 2| & \verb|Model 4| & \verb|Model 5| \\
\hline
\hline
P18+SH0ES &  1:300 & 1:100000 & 1:100000 \\
\hline
P18+DES+SH0ES &  1:2500 & 1:100000 & 1:1250 \\
\hline
P18+BAO+SH0ES &  1:14 & 1:100 & 1:20 \\
\hline
P18+SN+SH0ES &  1:300 & 1:500 & 1:500 \\
\hline
P18+BAO+SN+SH0ES &  1:16 & 1:50 & 1:20 \\
\hline
P18 &  1:500 & 1:4 & 1:1100 \\
\hline
P18+DES &  1:4 & 1:12 & 1:5 \\
\hline
P18+BAO &  {\rm fair} & {\rm fair} & {\rm fair} \\
\hline
P18+SN &  {\rm fair} & {\rm fair} & {\rm fair} \\
\hline
P18+BAO+SN &  {\rm fair} & {\rm fair} & {\rm fair} \\
\hline
\end{tabular}
\cprotect\caption{Betting odds against either $\Omega_{k}\neq 0$ (\verb|Model 4|) 
or $\omega_{k}\neq 0$ (\verb|Model 5|) in the data, respectively. 
For reference, betting odds against $\Omega_{k}\neq 0$ 
in K$\Lambda$CDM (\verb|Model 2|) are shown.}
\end{table}

Whereas the ultimate model comparison test must consider the overall model fitness to data, as quantified by the corresponding DIC values in Tables III and IV, it is sometimes useful to compare results for `anomaly' parameters, which should be identically zero in the reference model (flat $\Lambda$CDM). 
Significant evidence in the data for the addition of an `anomaly' parameter is indicative for underperformance of the reference model in comparison with the new modified models. The significance of statistical evidence for  such an anomaly parameter is usually reported in terms of percentiles or equivalently in terms of `betting odds' against a finite value for the parameter. 
We have chosen the latter option; the results are listed in Tables V -VI. 
For example, in Table V we report the corresponding results for the two fundamental parameters 
$\Omega_{k}$ and $\omega_{k}$, as well as for the derived parameter $\kappa$. It is clear 
from the table that $\kappa$ is in general a better diagnostic of the departures from the reference 
model than either $\Omega_{k}$ or $\omega_{k}$. It is also evident from Table VI that whereas 
K$\Lambda$CDM outperforms flat $\Lambda$CDM in virtually all dataset combinations considered 
in the present work when the SH0ES prior is imposed (open geometry), and also in the case of P18 and P18+DES even 
with the SH0ES prior excluded (closed geometry), \verb|Model 5| and \verb|Model 4|, respectively, do so even more significantly. We stress again that focusing on a single parameter might be misleading, given that the ultimate test is the `volume' of the likelihood 
function in parameter space: The smaller is the volume, the better is the fit to the data; indeed, this is 
gauged by the DIC values reported in Table III and IV. The general conclusions and trends deduced from Tables V-VI bode well with the results that are shown in Tables III and IV.

Triangle plots for selected cosmological parameters inferred from fitting \verb|Model 3| to various dataset combinations are shown in Figure 1. There is a clear correlation 
between $\kappa$ and both $\Omega_{b}h^{2}$, $\Omega_{\Lambda}$ and $\sigma_{8}$, and anti-correlation with both $\Omega_{c}h^{2}$, $\Omega_{k}$ and $\omega_{k}$. 
For $H_{0}$, $\Omega_{m}$ and $S_{8}$ both correlations and anti-correlations are  
seen, depending on the data combination used. If $S_{8}$ rather than $\sigma_{8}$ better captures the impact of growth of structure on the cosmological observables considered in this work, especially the DES dataset, then it is evident from Figure 1 that once $\kappa$ is allowed to freely vary, the data favors lower $S_{8}$ and higher $H_{0}$ values, even well and beyond those obtained with K$\Lambda$CDM and surely with flat $\Lambda$CDM. Specifically, these parameters inferred from fitting \verb|Model 3| to P18+DES results in $S_{8}=0.773\pm 0.017$ and $H_{0}=76.5\pm 3.0$ km/sec/Mpc at 68\% confidence level. For reference, when the same dataset is fitted with K$\Lambda$CDM (flat $\Lambda$CDM) 
we obtain $S_{8}=0.795\pm 0.016$ ($0.8018\pm 0.0066$) and $H_{0}=70.1\pm 1.7$ ($68.16\pm 0.48$) km/sec/Mpc.

\section{Discussion}

The standard GR-based {\it flat} $\Lambda$CDM model has been remarkably successful in 
describing a broad spectrum of cosmological phenomena probed with a wide variety of CMB projects, 
optical and IR telescopes, and extensive galaxy surveys, aimed at spectral and spatial mapping of the 
CMB, and tracing the large scale structure and evolution of the Universe. The K$\Lambda$CDM model 
is a better fit to all dataset combinations than $\Lambda$CDM when the SH0ES prior is assumed, whereas 
the two models perform equally well when the SH0ES prior is not adopted. 
However, when the models are contrasted only with the P18 dataset without the SH0ES prior, 
there is a strong preference for K$\Lambda$CDM. When the SH0ES prior is not included, the 
dataset combinations P18+BAO, P18+SN and P18+SN+BAO show no preference for 
K$\Lambda$CDM over the simpler flat $\Lambda$CDM model.
Consequently, if the SH0ES prior is ignored in the analysis the concordance model converges on flat $\Lambda$CDM by default. 

More generally, it is still relevant to ask whether the above two versions of the SM should be 
the only ones to consider in light of some theoretical uncertainties and given the complex nature 
of the various datasets. There are known to be a few longstanding cosmological anomalies that 
cannot be explained within either flat or K$\Lambda$CDM models. Each of these anomalies is 
normally parameterized by a single additional parameter over the basic parameters of $\Lambda$CDM 
or K$\Lambda$CDM models, effectively resulting in a better fit the current data. {\it The pressing 
question is whether $\Lambda$CDM is the `signal', or rather a `noise' in the vast space of possible phenomenological models of the Universe}. Bayesian analysis can indicate reasonably well which 
model is favored by the data given a certain set of priors. While statistical preference with no sound theoretical background may not provide (by itself) a sufficient basis for preferring one model over 
others, it is also true that the currently accepted cosmological model that relies on the well-established 
GR still requires invoking CDM and DE components. The latter is known to be anomalously small 
compared to theoretical predictions, let alone the fact that the true microphysics of CDM is unknown. 
Thus, framing the standard cosmological model within the well-established GR comes at a high price. Specifically, since the cosmological constant has the unique feature that it does not cluster there is no 
way to tell the difference between an anomaly in the Friedmann equation (in which case what seems 
like an accelerated expansion is not driven by a vacuum-like species) and a truly non-clustering form of 
energy within the GR-based model. In other words, the apparent acceleration might not be driven 
by a certain type of energy density -- it might just be an anomaly that is observed on the largest observable cosmological scales, certainly not on Galactic scales, and surely not in our solar system where GR has been 
reliably tested. Given this state of affairs, and especially the possibility that the DE parameter in 
the best-fit Hubble function represents an anomaly of GR on cosmological scales rather than a 
genuine non-clustering vacuum-like component of energy, it is of interest to explore the possibility 
that other anomalies are in play at the Friedmann equation level.

The unique spacetime metric that describes a maximally symmetric spatial hypersurface is the FRW 
metric with possibly a non-vanishing (constant) spatial curvature. The curvature parameter appears in 
both the Friedmann equation and geodesic equations, a single parameter that appears in both the t(z) 
and r(t) relations, and thus also in the r(z) relation. In this work we explore the consequences of adding 
an additional degree of freedom by allowing two independent curvature parameters in the t(z) and r(t) relations, as described in section III. We define a derived `anomaly' (the `curvature slip') parameter, 
$\kappa\equiv\omega_{k}-\Omega_{k}$ that quantifies the degree of mismatch between these two parameters, which we infer by determining the statistical significance of its departure from its canonical value of zero. The reference model, \verb|Model 1|, is $\Lambda$CDM, and \verb|Model 2| is K$\Lambda$CDM. Our modified models are \verb|Model 3|, in which both $\Omega_{k}$ and 
$\omega_{k}$ are free parameters; in \verb|Model 4| only $\Omega_{k}$ is a free parameter, 
whereas in \verb|Model 5| only $\omega_{k}$ is allowed to vary. Interestingly, if Eq. (9) is recast in 
the form $E(z)\equiv\sqrt{\Omega_{r}(1+z)^{4}+\Omega_{m}(1+z)^{3}+(\omega_{k}+\Omega_{\kappa})(1+z)^{2}+\Omega_{\Lambda}}$ where $\Omega_{\kappa}\equiv -\kappa$, then combined with Eq. (10) it could be {\it naively} interpreted as follows: At the {\it background level}, 
the model considered in this work looks like K$\Lambda$CDM with a dimensionless curvature 
$\omega_{k}$ sourced by the standard energy budget (radiation, NR matter, DE) and supplemented by 
an exotic source characterized by an effective EOS $-1/3$ and energy density in critical density units, 
$\Omega_{\kappa}$. However, to be considered as a material source the contribution of its 
perturbations must also be added to the analysis. In addition, inspection of 
Figure 1 leads to the conclusion that $\Omega_{\kappa}<0$. Indeed, this is 
akin to the interpretation that $\Omega_{\kappa}$ (or equivalently $\kappa$) represents an anomaly, 
much like DE could represent an anomaly rather than a genuine contribution to the cosmic energy budget.   

When the SH0ES prior is imposed K$\Lambda$CDM is clearly a better fit to the data than 
$\Lambda$CDM, as is evident from Table IV. \verb|Model 3| does much better with a DIC gain of 
$\gtrsim 20$ when neither BAO nor SN data are included; when these datasets are included, the 
gain reduces to $\sim 7-12$, still better than K$\Lambda$CDM. Remarkably, with just the P18+SH0ES 
data, \verb|Model 3| results in $\kappa\neq 0$ at very high statistical significance with betting odds 
worse than $1:10^{7}$ against $\kappa\neq 0$ (Table V). A similar conclusion applies to P18+DES+SH0ES and P18+BAO+SH0ES but with a lower statistical 
significance. \verb|Model 4| and \verb|Model 5| are also found to be statistically acceptable, 
though at a lower level of significance, as is apparent from the results listed in Tables V and VI.

For dataset combinations with no SH0ES prior, we find that none of 
\verb|Models 3|- \verb|5| 
provides a better fit to the data than does \verb|Model 2|. With these dataset combinations \verb|Model 3| 
does not provide a better fit over either \verb|Model 4| or \verb|Model 5|, except for the cases P18+DES 
or P18+SN, for which there is a strong evidence (supported by DIC gain $\gtrsim 7$) in favor of this 
model over both K$\Lambda$CDM and $\Lambda$CDM. We conclude that a model that allows for simultaneous variation of both $\Omega_{k}$ and $\omega_{k}$ is in general not warranted by data combinations that include BAO.   

Even though the above analysis is based on the addition of a phenomenologically-motivated parameter, 
it is nonetheless of interest to consider its possibly physical interpretations. As discussed above, the 
proposed models 
are better fits to current datasets than K$\Lambda$CDM and particularly $\Lambda$CDM, 
when the SH0ES prior is imposed. Since $H_{0}$ and $\Omega_{k}$ are correlated in 
K$\Lambda$CDM, then imposing the SH0ES prior drives $\Omega_{k}$ to positive values, i.e. towards preference for hyperbolic 
topology. The $H_{0}-\Omega_{k}$ correlation is intuitively clear: in a spatially closed 
Universe, i.e. $\Omega_{k}<0$, incoming light rays are bent inwards, the crossover time of a 
photon is longer, and consequently $H_{0}$ is smaller. Therefore, a higher value of $H_{0}$ results 
in a higher value of $\Omega_{k}$.  
A sufficiently high value of $H_{0}$ necessarily leads to $\Omega_{k}>0$, i.e., open Universe. 
Now, when both $\omega_{k}$ and $\Omega_{k}$ are allowed to simultaneously 
vary then even a flat geometry, $\omega_{k}\approx 0$ (that favors a higher 
$H_{0}$ than in closed space due to the above-mentioned correlation) is possible, but still have a very negative 
$\Omega_{k}$ that allows for a longer fly-time of incoming photons. (In the extreme case that $\Omega_{k}$ is sufficiently negative, i.e. the Universe is strongly closed, the fly-time could be 
infinite). By doing so, we can allow for a large $H_{0}$ to be still consistent with a very negative 
$\Omega_{k}$ without affecting the angular diameter distance and without having to compensate 
(by lowering the value of $H_{0}$) for the closed geometry simply because even though 
$\Omega_{k}$ can be significantly negative, $\omega_{k}$ can still satisfy $|\omega_{k}|\ll 1$.

\section{Summary}

With the ever-increasing precision of cosmological observations the challenge to $\Lambda$CDM 
is correspondingly higher. As the number and statistical significance of SM anomalies increase, so is the 
interest in alternative models. The SM is based on our best-tested theory of gravitation, GR, 
that withstood the test of time. However, GR is well-tested only within our solar system, whereas 
on galactic and supergalactic scales it has fared poorly as manifested by the need to invoke the 
existence of CDM and DE. Whereas on the largest cosmological scales these two mass-energy 
forms are described by only two new parameters, and therefore add only a very slight complexity 
to the model, this is far from being the case on galactic scales where CDM profiles have to be 
fit separately for each and every galaxy or cluster of galaxies. As argued here -- and to the best 
of our knowledge nowhere else in the context of the present work -- what is known as `Dark Energy' (assuming its simplest form -- a cosmological constant) could by itself represent a gravitational anomaly on the largest cosmological scales, rather than a genuine form of energy.

The P18 data strongly favors a spatially closed space, but  only with a relatively low value of 
$H_{0}$ which is at odds with local measurements of $H_{0}$. Other datasets, e.g. P18+DES even 
favor an open Universe; only P18+SN supports the closed Universe model but at very weak 
$\sim 1\sigma$ confidence level. The BAO dataset has played a (somewhat surprisingly) significant 
role in the acceptance of flat $\Lambda$CDM as the SM. It would be remarkable if indeed the 
Universe with its (post-inflation) continuous growth of structure has no appreciable departure 
from global spatial flatness -- yet this is what the current varied datasets, interpreted within 
the framework of the GR-based SM, seem to suggest. In the present work we relaxed the `stiffness' 
of K$\Lambda$CDM to test for the possibility that a less `rigid' model better fits the data. While 
doing so we found that especially when local measurements of $H_{0}$ are  adopted as a prior 
(in the statistical analysis), this generalized model better fits the data than K$\Lambda$CDM and 
all the more better than $\Lambda$CDM. This new anomaly, the `curvature slip' (gauged by the departure of a 
newly introduced $\kappa$ parameter from its vanishing canonical value), is yet one more in a growing list of SM anomalies that enhance 
the need for either new physics or more nuanced analyses 
within the realm of standard physics. 

\section*{Acknowledgements}
This research has been supported by a grant
from the Joan and Irwin Jacobs donor-advised fund at the
JCF (San Diego, CA).\\

\end{document}